\documentclass[fleqn]{article}
\usepackage{espcrc2}
\usepackage{graphicx}
\usepackage[figuresright]{rotating}
\def\Jvol<#1,#2,#3>{#1}
\def\Jpage<#1,#2,#3>{#2}
\def\Jyear<#1,#2,#3>{#3}
\newcommand\journal[1]{\textbf{\Jvol<#1>}, \Jpage<#1> (\Jyear<#1>)}
\newcommand\PRL[1]{Phys.\ Rev.\ Lett.\ \journal{#1}}
\newcommand\PRC[1]{Phys.\ Rev.\ C \journal{#1}}
\newcommand\PRD[1]{Phys.\ Rev.\ D \journal{#1}}
\newcommand\PLB[1]{Phys.\ Lett.\ B \journal{#1}}

\newcommand\NPA[1]
{Nucl.\ Phys.\ \textbf{A\Jvol<#1>}, \Jpage<#1> (\Jyear<#1>)}
\newcommand\NPB[1]
{Nucl.\ Phys.\ \textbf{C\Jvol<#1>}, \Jpage<#1> (\Jyear<#1>)}
\newcommand\NPBS[1]
{Nucl.\ Phys. (Proc.Suppl.)\ \textbf{B\Jvol<#1>}, \Jpage<#1> (\Jyear<#1>)}

\newcommand\NPS[1]
{Nucl.\ Phys. (Proc.Suppl.)\ \textbf{\Jvol<#1>}, \Jpage<#1> (\Jyear<#1>)}

\newcommand\PTP[1]{Prog.\ Theor.\ Phys.\ \journal{#1}}

\newcommand\ibid[1]{\textit{ibid.}\ \journal{#1}}

\newcommand{\be}{\begin{equation}}
\newcommand{\ee}{\end{equation}}
\newcommand{\hmu}{\hat{\mu}}

\title{Properties of hadron screening masses at small baryonic density}

\author{
QCD-TARO Collaboration:
Irina Pushkina\address{School of Biosphere Science, Hiroshima
University, Higashi-Hiroshima 739-8526, Japan}
\thanks{E-mail: irina@riise.hiroshima-u.ac.jp},
Philippe de Forcrand\address{
Institut f\"ur Theoretische Physik,
ETH-H\"onggerberg, CH-8093 Z\"urich, Switzerland, and
Theory Division, CERN, CH-1211 Geneva 23, Switzerland}, 
Margarita Garcia Perez\address{
Instituto de F\'{\i}sica Te\'orica, Universidad Aut\'onoma de
Madrid, ~Cantoblanco, 28049 Madrid, Spain},
Seyong Kim\address{
University of Wales, Swansea, and
Department of Physics, Sejong University, Seoul 143-747, Korea},
Hideo Matsufuru\address{
Computing Research Center, High Energy Accelerator Research
Organization (KEK), ~Tsukuba 305-0801, Japan},       
Atsushi Nakamura\address{
RIISE, Hiroshima University, Higashi-Hiroshima 739-8521, Japan},
Ion-Olimpiu Stamatescu\address{
Institut f\"ur Theoretische Physik, Universit\"at
Heidelberg,  D-69120 Heidelberg, Germany,
and FEST, Schmeilweg 5, D-69118 Heidelberg, Germany},
Tetsuya Takaishi\address{
Hiroshima University of Economics, Hiroshima 731-0192, Japan}
and
Takashi Umeda\address{
Yukawa Institute for Theoretical Physics, Kyoto University
Kyoto 606-8502, Japan}
}

\begin{document}

\begin{abstract}
The properties of hadron screening masses around the deconfinement phase transition at finite baryonic density can be studied by evaluating the Taylor coefficients with respect to the iso-scalar ($\mu_S=\mu_u=\mu_d$) and iso-vector ($\mu_V=\mu_u=-\mu_d$) chemical potentials, where $\mu_u$ and $\mu_d$ are $u$ and $d$ quark chemical potentials.
We simulate 2-flavour lattice QCD with staggered fermions on a $12^2\times 24\times 6$ lattice with  $ma = 0.05$ and $0.10$ and report investigations of nucleon, pseudo-scalar and vector mesons.
We present new, strong evidence that in the confining phase, the screening masses at $\mu=0$ have significant temperature dependence, but the effect of $\mu_S$ is very small. 
Above the critical temperature, the second derivative terms of mesons rapidly increase as contrasted to the case of baryon.
We also study the responses of the screening masses to an iso-vector chemical potential and discuss some of the issues related to the properties of hadron masses at finite $\mu$.
\end{abstract}

\maketitle

\section{Introduction}

There have been extensive studies of properties of hadron masses at finite temperature using phenomenological models \cite{models,FuHaLee} and lattice QCD simulations \cite{L-FiniteT-Screening,L-FiniteT-Pole,L-FiniteT-Jpsi}, while little is known about their behavior in the medium. 
Although many model studies and simulations of two-color QCD produced interesting outcomes \cite{ModelFiniteMu}, no lattice QCD study of hadron pole masses at finite baryon density has yet been performed.
This is due to the difficulty in performing numerical simulations with a complex fermion determinant.
See Refs.~\cite{MNNT,MariaReview}.
  
The chemical potentials of $u$ and $d$ flavors can be combined into  iso-scalar and iso-vector (also called isospin) chemical potentials as $\hmu_S = \hmu_u = \hmu_d$ and $\hmu_V = \hmu_u = -\hmu_d$, respectively.
In this letter we study the response of hadron screening masses with respect to both $\hmu_S$ and $\hmu_V$.
In addition to the pseudo-scalar ($\pi^+(u\bar{d})$) and vector mesons, we also study the chemical potential response of the nucleon mass (n($udd$)).
To our knowledge, this is the first attempt to investigate baryon masses at finite baryonic density in SU(3) lattice QCD.

Our analysis focuses on the behavior in the vicinity of zero chemical potential $\mu=0$.
This is a region which currently running relativistic heavy-ion collision experiments at RHIC explore. 
In addition to the usual, baryonic (iso-scalar) chemical potential $\mu_S$, we also present data for the iso-vector chemical potential $\mu_V$ case.
Incident nuclei in high energy heavy ion collisions are usually neutron rich, and therefore the $\mu_V$ dependence reported here may be observed in these experiments.

Using a Taylor expansion method at fixed temperature $T$, coupling $\beta$ and bare quark masses, the screening mass is expanded at $\mu=0$ in the form,
\begin{eqnarray}\nonumber
\lefteqn{\frac{M(\mu)}{T} = \frac{M}{T}\Bigg|_{\mu=0} +
\left(\frac{\mu}{T}\right)\frac{\partial M}{\partial\mu}\Bigg|_{\mu=0}} \\
& & +\left(\frac{\mu}{T}\right)^2\frac{T}{2}
\frac{\partial^2 M}{\partial\mu^2}\Bigg|_{\mu=0} +
O\left[\left(\frac{\mu}{T}\right)^3\right].
\label{Eq:Taylor}
\end{eqnarray}
This allows to perform numerical simulations with standard methods: we determine the Taylor coefficients at $\mu=0$, by measuring hadron propagators and their derivatives in $\mu=0$ lattice QCD calculations~\cite{QCD-TARO}.

\section{Lattice setup}\label{lattice}

We study effects of the chemical potential on hadron screening masses with staggered fermions. 
The numerical simulations are carried out on a lattice of size $12^2\times 24\times 6$, at eight values of $\beta$ from 5.30 to 5.65, with quark masses $ma = 0.05$ and $0.10$.
Preliminary results were presented in Ref.~\cite{QCD-TARO,PrRes}. 
Earlier investigations were limited to smaller lattices $16\times 8^2\times 4$ and small number of $\beta$ values.
We report here new calculations around the phase transition, using lattices with a larger spatial volume and temporal extension $N_t=6$, which provides us with data nearer to the continuum limit.

The R-algorithm was used to generate 1000 configurations each separated by 10 unit-length trajectories, with the molecular dynamics step size set to 0.02.
Anti-periodic (periodic) boundary conditions were imposed in temporal (spatial) directions on the fermion fields.

We deduced the scale from previous estimates of lattice spacings $a$ for various values of $\beta$ and quark masses determined by Tamhankar and Gottlieb~\cite{Somali}. 
In the present analysis the lattice spacing $a$ varies between 0.09 and 0.26 fm.

The critical value of the coupling ($\beta_c$) is estimated from the peak of the Polyakov line susceptibility, giving $\beta_c = 5.499$ for $ma = 0.05$ and $\beta_c = 5.568$ for $ma = 0.10$. 
The values of $\beta$ studied here are listed in Table 1 together with a rough estimate of $T/T_c$.

To improve the ground state signal, we introduced a "corner" wall source. 
This source is not gauge invariant, so we fixed the link variables in the spatial slice $z$ to the "Coulomb" gauge for each configuration. 

The Taylor coefficients in Eq.~(\ref{Eq:Taylor}) are obtained by taking derivatives of hadron correlators, for which we need the traces of various fermionic operators.
To evaluate these traces, we used 200 "noise" vectors of complex random numbers.

\section{Hadron screening masses}\label{hadron}

\begin{figure}[htb]
 \begin{center}
 \includegraphics[width=18pc]{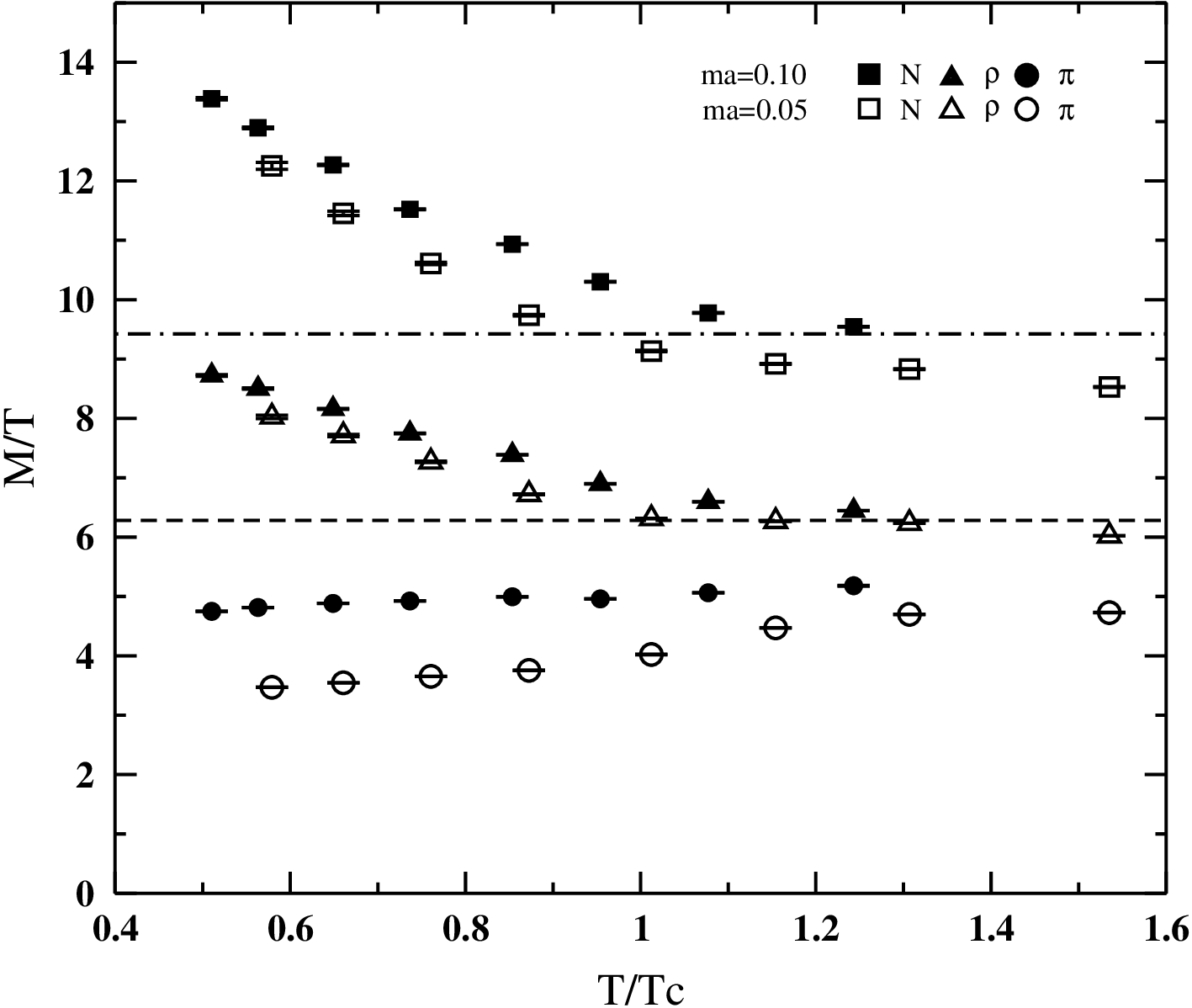}
  \caption{The screening masses of pseudo-scalar (PS), vector (V) mesons and nucleon (N), divided by the temperature $T$.
The dashed (dot-dashed) line represents the free-quark limit $2\pi$ ($3\pi$).
Open symbols correspond to the quark mass $ma = 0.05$ and filled ones to $ma = 0.10$.}
  \label{fig1}
 \end{center}
 \end{figure}

Figure~\ref{fig1} shows the temperature dependence of hadron screening masses obtained from fitting the correlators to a double-exponential form, \footnote{The sign of the second term in each bracket is opposite to the standard formula~\cite{fit} because of the periodic boundary conditions imposed in the spatial directions.}

\begin{eqnarray}\nonumber
C_\pi(z) & = & C_1\,\left(e^{-\hat{m}_1\,\hat{z}} +
                          e^{-\hat{m}_1\,(N_z - \hat{z})}\right) \\ \nonumber
         & + & C_2\,\left(e^{-\hat{m}_2\,\hat{z}} + 
                          e^{-\hat{m}_2\,(N_z - \hat{z})}\right),\\ \label{cor}
C_\rho(z) & = & C_1^\prime\,\left(e^{-\hat{m}_1^\prime\,\hat{z}} +  
                            e^{-\hat{m}_1^\prime\,(N_z - \hat{z})}\right) \\ \nonumber
          & + & C_2^\prime\,(-1)^z\,\left(e^{-\hat{m}_2^\prime\,\hat{z}} +
                            e^{-\hat{m}_2^\prime\,(N_z - \hat{z})}\right),\\ \nonumber
C_N(z) & = & C_1^{\prime\prime}\,\left(e^{-\hat{m}_1^{\prime\prime}\,\hat{z}} + 
              (-1)^z\,e^{-\hat{m}_1^{\prime\prime}\,(N_z - \hat{z})}\right) \\ \nonumber
       & + & C_2^{\prime\prime}\,\left((-1)^z\,e^{-\hat{m}_2^{\prime\prime}\,\hat{z}} +
                              e^{-\hat{m}_2^{\prime\prime}\,(N_z - \hat{z})}\right),
\end{eqnarray}
where $N_z$ is the length of the lattice in the $z$-direction. 
We write the correlators in terms of dimensionless lattice variables, by scaling $m_1$, $m_2$ and $z$ according to their canonical dimensions.

We start our analysis with the hadron screening mass which is obtained as a a fitting parameter in each channel. Its value is the zeroth order term in the Taylor expansion of the hadron mass, i.e., $M(\mu=0)$.
For the pseudo-scalar meson, we do not consider a contribution with the alternating sign. In this case, the two-particle fit includes an excited state with the same quantum numbers. 
For the vector meson and nucleon the two-particle fit includes one particle of each parity ~\cite{fit}. We have observed that a single-exponential fit is not good enough to extract reliable results in the deconfinement phase, where parity partners are expected to be degenerate.

The $\mu=0$ screening masses divided by $T$ are shown in Fig.~\ref{fig1}. 
As expected, the pseudo-scalar and vector mesons become degenerate above $T_c$ because of chiral symmetry restoration. With increasing temperature, the screening masses of mesons (baryon) approach $2\pi T$ ($3\pi T$) which corresponds to free quarks having the minimal momentum $\pi T$~\cite{Learman01}.
The vector meson screening mass approaches this asymptotic value at a surprisingly low temperature.

\begin{table}[htb]
\begin{center}
 \caption{Values of $\beta$ studied here and corresponding temperatures.}
\begin{tabular}{c||c|c}\hline
          & $ma = 0.05$ & $ma = 0.10$  \\
 $\beta$  & $T/T_c$     & $T/T_c$      \\ \hline
   5.30   &  0.58       &  0.51        \\
   5.35   &  0.66       &  0.56        \\
   5.40   &  0.76       &  0.64        \\
   5.45   &  0.87       &  0.74        \\
   5.50   &  1.01       &  0.85        \\
   5.55   &  1.15       &  0.95        \\
   5.60   &  1.31       &  1.08        \\
   5.65   &  1.54       &  1.24        \\ \hline
\end{tabular}
\end{center}
\end{table}

\subsection{The pseudo-scalar meson}\label{pion}

The first derivatives of both the pseudo-scalar and the vector meson
correlators with respect to the chemical potential vanish. 
In the case of the iso-scalar chemical potential, this can be easily seen from the symmetry $\mu_S \to -\mu_S$ of the correlator.  See Ref.~\cite{QCD-TARO,PrRes}.
The first derivatives with respect to the iso-vector chemical potential are numerically also consistent with zero.

\begin{figure}[htb]
 \begin{center}
 \includegraphics[width=18pc]{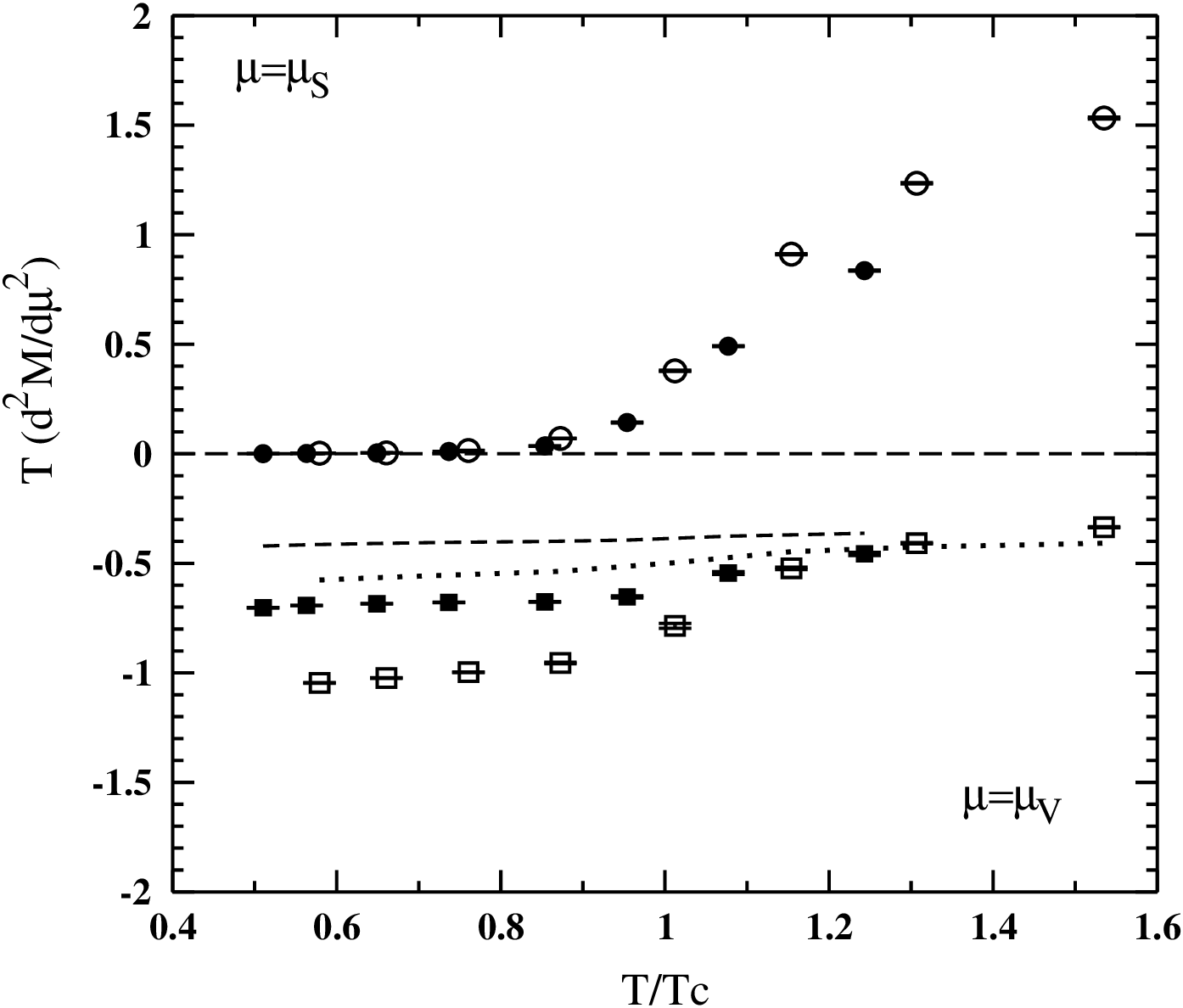}
  \caption{The second order response of pseudo-scalar meson. 
The open symbols correspond to the smaller quark mass $ma = 0.05$, the filled ones to $ma = 0.10$.
The dotted (dashed) line stands for $-2T/M_{PS}$, where $M_{PS}$ is the pseudo-scalar meson mass at $ma=0.05$ ($ma=0.10$) obtained from Fig.1.}
  \label{fig2}
 \end{center}
 \end{figure}

The second order responses of the pseudo-scalar meson correlator are shown in Fig.~\ref{fig2}. In the confined phase, the response to $\mu_S$ is small: the pion is protected from acquiring a large mass because chiral symmetry is broken.
In the deconfinement phase, the response increases with temperature and therefore the pseudo-scalar meson becomes heavier when $\mu_S$ is finite.
In the same figure we show data for two different quark masses $ma=0.10$ and $ma=0.05$. 
With decreasing quark mass, the $\mu_S$ response increases for $T>T_c$. 

The response with respect to $\mu_V$ shows a very different behavior.
In the confinement phase it is negative, while above $T_c$ it approaches zero.
If we insert the values into Eq.~(\ref{Eq:Taylor}), neglecting the $O[(\mu/T)^3]$ terms, the mass becomes zero around $\mu_V \sim M_\pi$, which is the expected critical value
of the iso-vector chemical potential where a pion condensate forms \cite{SonStephanov}.
This indicates that the pion mass dependence on $\mu_V$ is rather well described by our truncated quadratic Taylor expansion in the whole confining phase.
Equivalently, the second-order response can be compared with $-2T/M_{PS}$ (Fig.~\ref{fig2}), i.e. the value it would take if the quadratic ansatz were exact.

\subsection{The vector meson}\label{rho}

\begin{figure}[htb]
 \begin{center}
 \includegraphics[width=18pc]{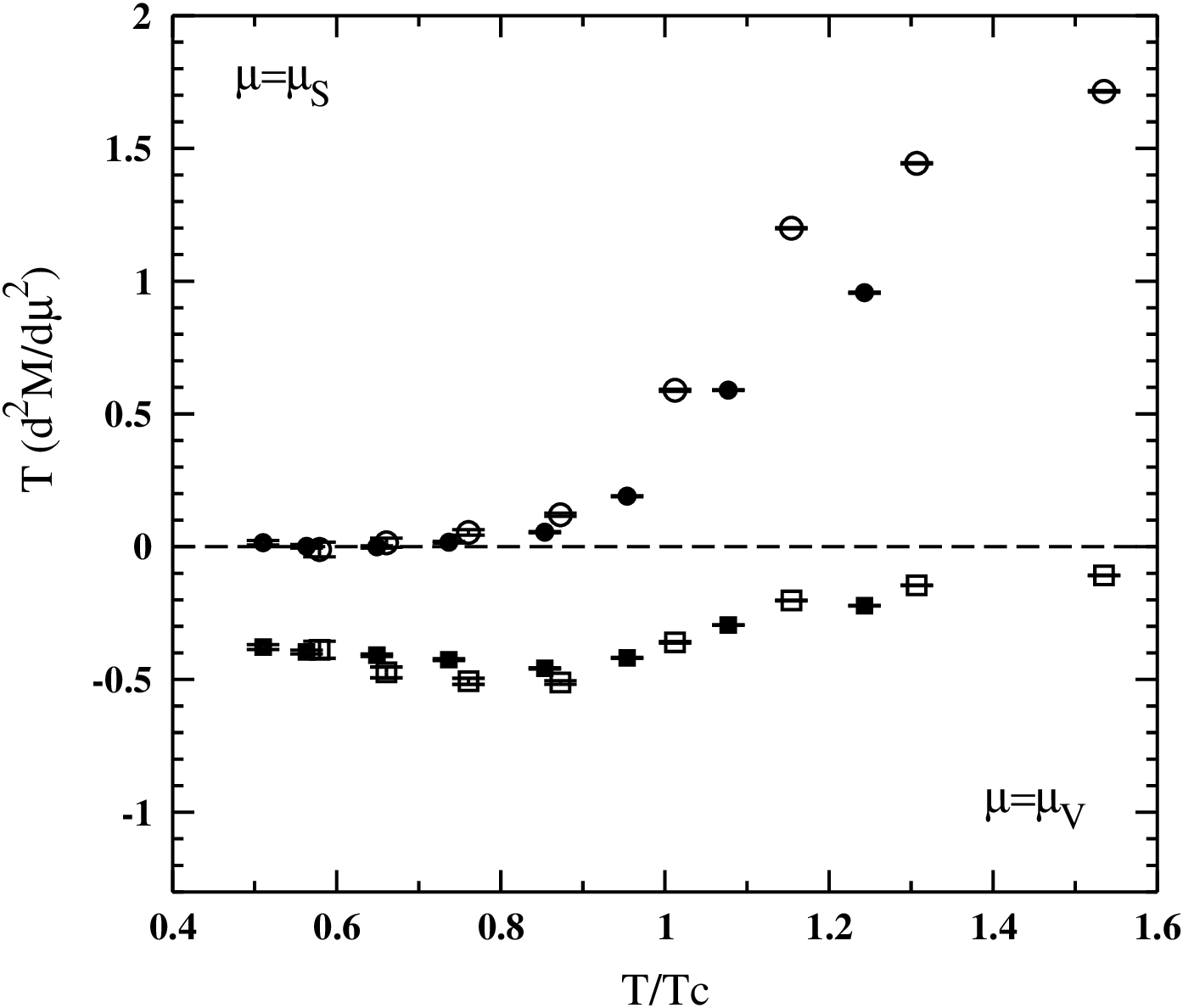}
  \caption{The second order response of the vector meson. 
Open symbols correspond to the quark mass $ma = 0.05$ and filled ones to $ma = 0.10$.}
  \label{fig3}
 \end{center}
 \end{figure}

As shown in Fig.~\ref{fig3}, the situation with the vector meson is quite similar to the case of the pseudo-scalar meson described above. 
In the confinement phase, the response to $\mu_S$ is almost zero, and increases in the deconfinement region. 
This indicates that the vector meson screening mass above $T_c$ increases in the medium, assuming it still represents relevant degrees of freedom there.
 
The second derivative with respect to $\mu_V$ below $T_c$ shows similar behavior as the pseudo-scalar case.
Above $T_c$ it approaches zero quickly and has little quark mass dependence.

\subsection{Baryon (nucleon $udd$)}\label{nucleon}

Unlike the case of mesons, the first order response of the nucleon does not vanish for either $\mu_S$ or $\mu_V$ (see Fig.~\ref{fig4}).

The derivative with respect to the iso-scalar chemical potential is positive and close to zero at $T<T_c$.
It increases quickly in the deconfinement phase, and more so for smaller quark mass.
The response to $\mu_V$ is again negative. 

As for the second order response (Fig.~\ref{fig5}), it is positive under the influence of $\mu_V$ and negative in the case of $\mu_S$. 
The difference between the two quark mass cases is very small till $1.1 T_c$.

The second order term in the Taylor expansion is of the same order as the first one, and gives a contribution of opposite sign. 
Consequently, the nucleon mass is less sensitive to chemical potential than meson masses.

\section{Concluding Remarks}\label{conclusion}

We have analyzed the behavior of the first and second order derivatives of the meson and baryon screening masses with respect to the chemical potential for two quark masses $ma=0.05$ and $ma=0.10$. 

 \begin{figure}[htb]
 \begin{center}
 \includegraphics[width=18pc]{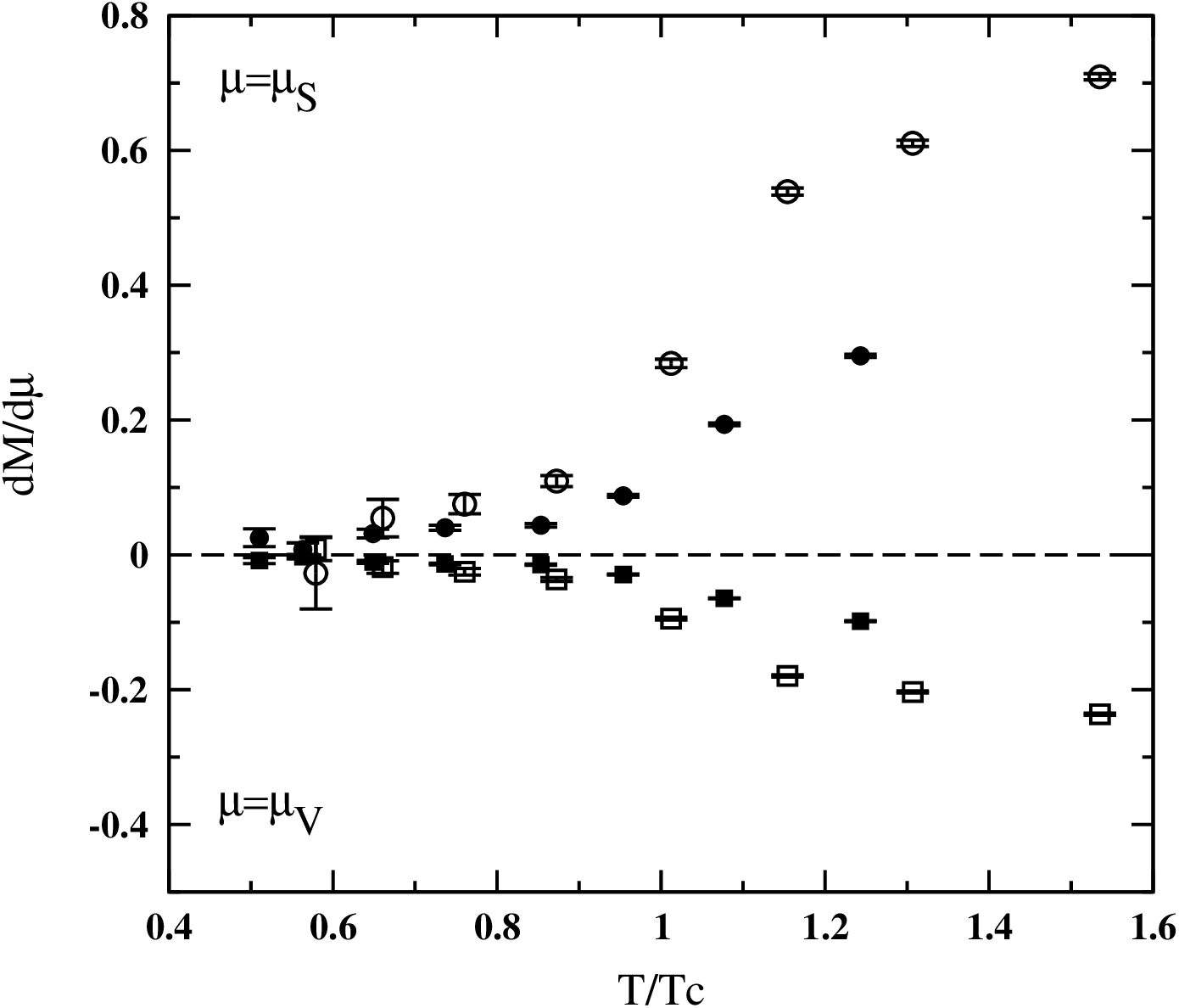}
  \caption{First order response of the nucleon (neutron).
Open (filled) symbols correspond to the quark mass $ma = 0.05$ ($ma = 0.10$).} 
  \label{fig4}
 \end{center}
 \end{figure}
%
 \begin{figure}[htb]
 \begin{center}
 \includegraphics[width=18pc]{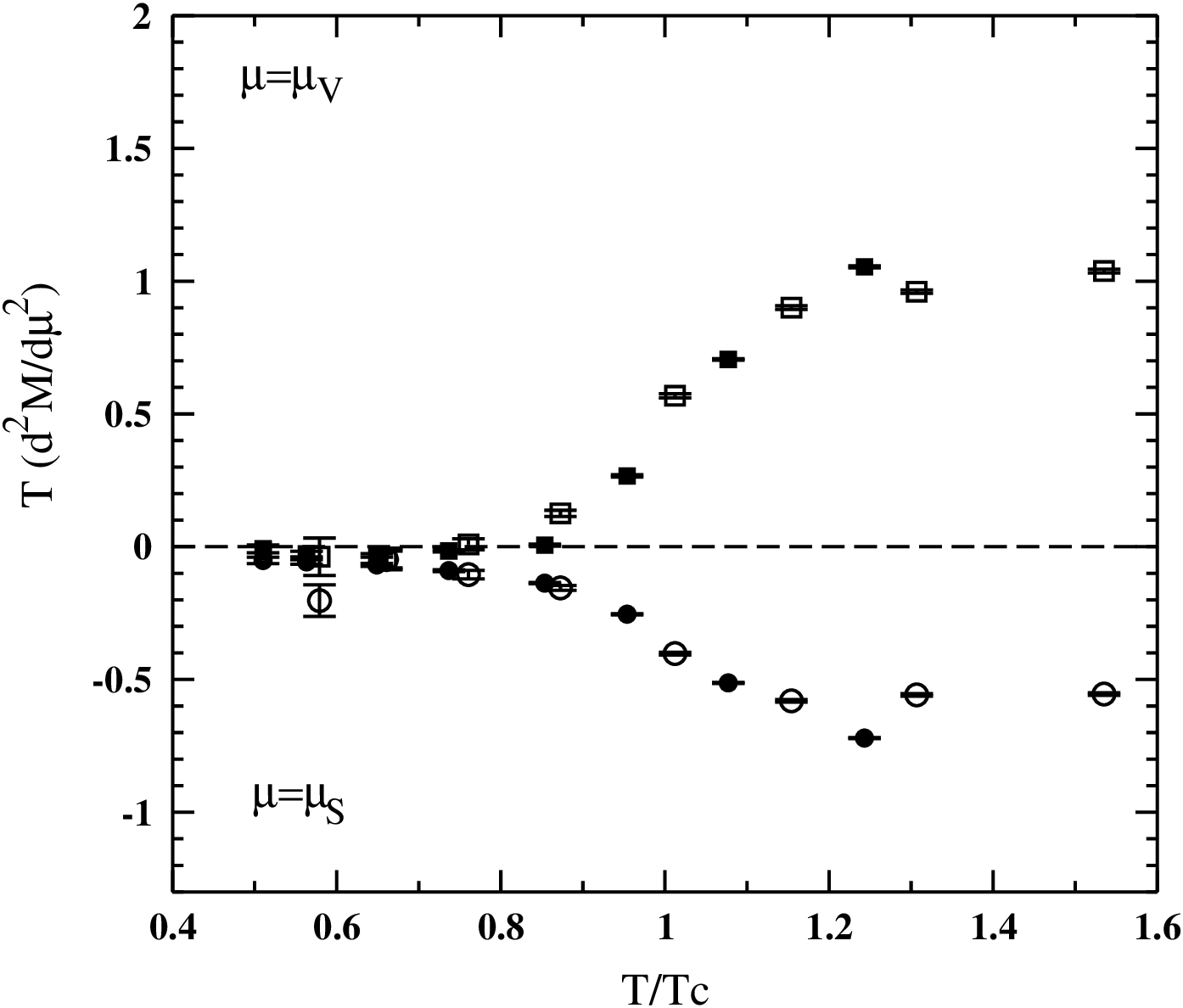}
  \caption{Second order response of the nucleon (neutron).
Open (filled) symbols correspond to the quark mass $ma = 0.05$ ($ma = 0.10$).}
  \label{fig5}
 \end{center}
 \end{figure}

Our study shows that in the vicinity of zero iso-scalar chemical potential, the hadron screening mass increases as function of both temperature and density.
At high temperature, the second order derivative with respect to $\mu_S$ increases and that w.r.t. $\mu_V$ approaches zero for both pseudo-scalar and vector meson cases.
At very high temperature the meson propagators should be superposition of free quarks. 
If we introduce an imaginary chemical potential $\mu_I$, in momentum space they are written schematically as

\begin{eqnarray}
\sum_{p^{(1)},p^{(2)}}\left(\gamma_0 (\sin(p_0^{(1)}+\mu_{Iu})+\vec{\gamma}\sin\vec{p}^{(1)}\right)^{\dagger -1}\nonumber \\
\Gamma\left(\gamma_0 \sin(p_0^{(2)}+\mu_{Id})+\vec{\gamma}\sin\vec{p}^{(2)}\right)^{-1}
\Gamma^{\dagger},
\end{eqnarray}
where $p_0 = n\pi T$ with $n = \pm 1, \pm 2, \cdots$.
The baryon propagator is constructed in the same manner.

Such a "meson" propagator does not of course have a pole, but its behavior gives us a hint to understand what happens at finite chemical potential.
In case of the iso-scalar chemical potential, the lowest contribution to the damping in the $z$-direction  comes from $|-\pi T+\mu_I|+|-(-\pi T+\mu_I)| \sim 2(\pi T-\mu_I)$ if $\mu_I$ is positive and smaller than $2\pi T$.
For the iso-vector case, we have $|\pi T-\mu_I|+|-\pi T-\mu_I| \sim 2\pi T$.
Therefore when $T$ increases above $T_c$, the meson mass reaches $2\pi T $ showing little dependence on the imaginary iso-vector chemical potential, while with the iso-scalar chemical potential, the mass decreases and may behave as $M(0)-C\mu_I^2$ with $C$ positive.
If we now go back to a real chemical potential, $M(0)+C\mu^2$, i.e., the second derivative with respect to $\mu_S$ is positive, while that w.r.t. $\mu_V$ approaches zero as seen in Figs.\ref{fig2} and \ref{fig3}. 

The second response of the vector meson with respect to $\mu_S$ above the phase transition temperature is similar to that of the pseudo-scalar meson.
This indicates that when the chiral symmetry is restored, not only the masses of $\pi$ and $\rho$, but also their response to the chemical potential is similar.

Although the quark mass dependence below $T_c$ is different, the gross features of the response to $\mu_V$ of the $\rho$ meson are similar to that of the $\pi$.
This may indicate an interesting phenomena, i.e., in an experimental environment where the number of $u$ quarks is not equal to that of $d$ quarks, the vector meson mass may shift.

There have been many analyses of the finite iso-vector chemical potential system by NJL model~\cite{Toublan1,Choe02}, chiral lagrangians~\cite{SonStephanov}, random matrix model~\cite{Toublan2}, the strong coupling expansion~\cite{Nishida03} and lattice simulations~\cite{Gupta02,Sinclair04,Sasai03,Takaishi03,Takaishi04}.
Son and Stephanov first pointed out that at $\mu_V=m_\pi$ and $T=0$, there is a phase transition into a pion condensation phase where $\pi$ becomes massless \cite{SonStephanov}.
According to these analyses, the leading iso-vector chemical potential effect of the pion pole mass is a linear in $\mu_V$.
Our simulations give an insight into the region of finite temperatures where we observe that the first response of the screening mass w.r.t. $\mu_V$ vanishes.
 
In Ref.~\cite{SonStephanov00}, QCD is studied at finite $\mu_V$ in the low and high $\mu_V$ limits. The calculation of the neutron mass is performed using baryon chiral perturbation theory.

\begin{equation}
M = M(\mu=0) - \frac{\mu_V}{2}\cos(\alpha),
\end{equation}
where $\cos\alpha=1$ for $\mu_V<m_\pi$ and $\cos\alpha=m_\pi^2/\mu_V^2$ for $\mu_V \ge m_\pi$.
This may be consistent with the negative first derivative obtained here.

The value of the first response of the nucleon with respect to the iso-scalar chemical potential (Fig.~\ref{fig4}) is very small but positive.
If this is definitely non-zero, it means that at low but positive baryon density environment, the nucleon screening mass is heavier, i.e., the baryon exchange mechanism may
be suppressed. 

The present simulation was performed with two fixed values of $ma=$0.10 and 0.05.
In order to compare with experimental data, we should extrapolate them to the chiral limit $ma \rightarrow 0$. 
We plan to study the screening masses at finite $T$ and $\mu$ along a line of constant physics corresponding to the real (two flavor) world.

\bigskip
\noindent
{\bf Acknowledgment}
The simulation has been done on Hitachi SR8000 at KEK (High Energy Accelerator Research Organization) and SR8000 at Hiroshima University. 
Large data transformation was carried out through the SuperSINET. 
This work is partially supported by Grant-in-Aide for Scientific Research by Monbu-Kagaku-sho (No.\ 11440080, 12554008, 13135216, 14540263) and by Japanese Society for Promotion of Science (T.~Umeda).
S.K. is supported by KISTEP grant M6-0404-01-001 in part.

\end{document}